\documentclass[graybox]{svmult}

\usepackage{mathptmx}       
\usepackage{helvet}         
\usepackage{courier}        
\usepackage{type1cm}        
\usepackage{makeidx}         
\usepackage{graphicx}        
\usepackage{multicol}        
\usepackage[bottom]{footmisc}

\makeindex             

\begin{document}

\title*{Nonlocal model for the turbulent fluxes due to thermal convection in rectilinear shearing flow}
\titlerunning{Nonlocal model for the turbulent fluxes} 
\author{R. Smolec, G. Houdek and D.O. Gough}
\institute{R. Smolec and G. Houdek \at Institute of Astronomy, University of Vienna, Austria, \email{radek.smolec@univie.ac.at}
\and D.O. Gough \at Institute of Astronomy and Department of Applied Mathematics and Theoretical Physics, University of Cambridge, UK}
%
%
\maketitle

\abstract*{We revisit a phenomenological description of turbulent thermal convection along the lines proposed by Gough \cite{DG77a} in which eddies grow solely by extracting energy from the unstably stratified mean state and are subsequently destroyed by internal shear instability. This work is part of an ongoing investigation for finding a procedure to calculate the turbulent fluxes of heat and momentum in the presence of a shearing background flow in stars.}
\abstract{We revisit a phenomenological description of turbulent thermal convection along the lines proposed by Gough \cite{DG77a} in which eddies grow solely by extracting energy from the unstably stratified mean state and are subsequently destroyed by internal shear instability. This work is part of an ongoing investigation for finding a procedure to calculate the turbulent fluxes of heat and momentum in the presence of a shearing background flow in stars.}

\smallskip
We describe the first steps towards a generalization of the time-dependent formulation of the mixing-length approach for radial pulsation, proposed by Gough \cite{DG77a}, to nonradially pulsating stars. To this end, we incorporate a treatment of the influence of a shearing background flow. In order to test and calibrate the formalism, we first compare its predictions with results of direct numerical simulation (DNS) of Rayleigh-B\'enard convection in air (Prandtl number, $\sigma=0.71$) with a strongly shearing background flow \cite{DM88}. We consider a plane-parallel layer of fluid confined between rigid horizontal perfectly conducting boundaries of infinite extent at fixed temperatures, the lower being hotter than the upper by $\Delta T$. The upper boundary moves horizontally with constant velocity $\Delta U$ relative to the lower boundary, and at first we assume, in accordance with the Boussinesq approximation, that the shear, $E$, in the mean flow does not vary over the scale of an eddy. Previously we considered the local model only \cite{SHG}, which we extend here to account for nonlocal effects.

Following \cite{GH01} we solve the linearized dimensionless equations describing the dynamics in a statistically stationary flow of a viscous Boussinesq fluid confined between two horizontal planes. Linearized modes of convection are obtained by expansion about $E:={\rm d}_zU=0$ (${\rm d}_z\equiv{\rm d}/{\rm d}z$) to second order in $E$. The resulting expressions for the eigenfunctions of the fluctuating temperature, $T'$, and turbulent velocity field, $u_i=(u,v,w)$, are used to compute the turbulent fluxes of heat, $F_{\rm c}\propto\overline{wT'}$, and momentum (Reynolds stresses), $\overline{u_iu_j}$, in the manner of \cite{DG77b}. A horizontal bar denotes the average over the horizontal plane. The local expressions are next averaged in the manner proposed by \cite{DG77b} to account for nonlocal effects: the eddies of vertical extent $\ell$  centred at $z_0$ sample the temperature gradient over the spatial region $(z_0-\ell/2,\ z_0+\ell/2)$, and the nonlocal fluxes, $\langle F_{\rm c}\rangle$ and $\langle\overline{u_iu_j}\rangle$,  are determined from the contributions of all eddies centred between $z_0-\ell/2$ and $z_0+\ell/2$. These fluxes enter the conservation equations for heat, $N=-{\rm d}_z\overline{T}+\langle F_{\rm c}\rangle$, and momentum, $S={\rm d}_z\overline{U}-\langle\overline{uw}\rangle/\sigma$, which enable one to compute the mean temperature and velocity profiles, $\overline{T}(z)$ and $\overline{U}(z)$, across fluid layer. The total dimensionless heat flux (Nusselt number), $N$, and total stress, $S$, are eigenvalues of the system. The diffusion approximation for the radiative flux is used. 

\begin{figure}[t]
\includegraphics[scale=.45]{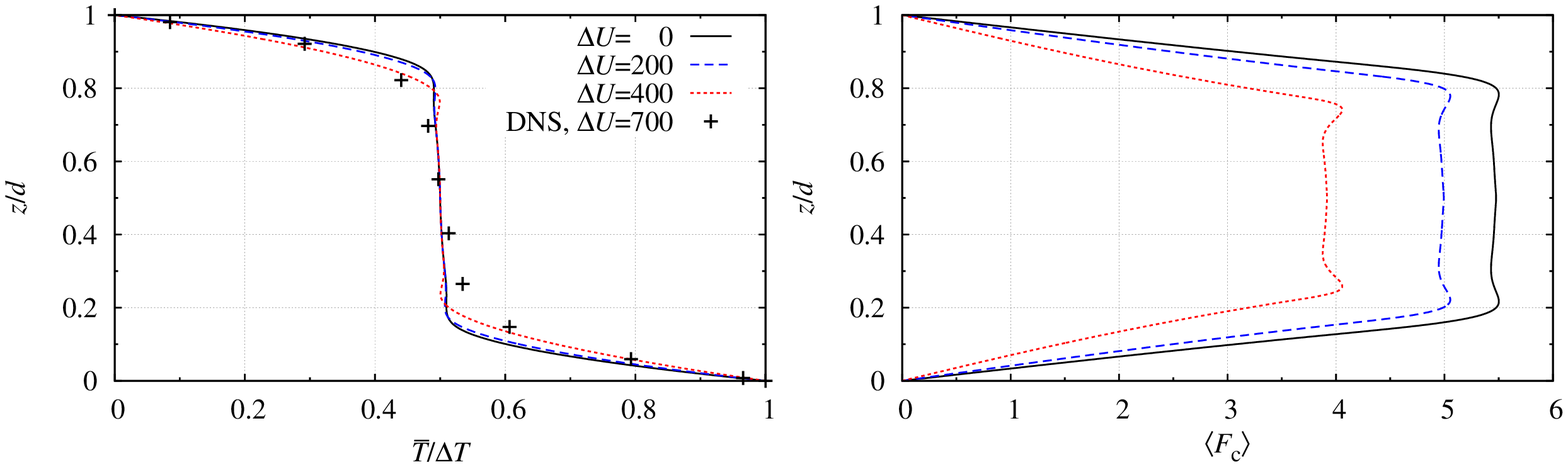} 
\caption{Mean temperature profile  $\overline{T}/\Delta T$ compared with DNS data (left panel) and dimensionless convective heat flux $\langle F_{\rm c}\rangle$ (right panel) as a function of height $z$ for different $\Delta U$ values.}
\label{fig1}
\end{figure}

In Fig.~\ref{fig1} (left panel) we compare the mean temperature profile with the DNS data. Agreement of the functional form of the mean temperature profile between our model results and DNS data is obtained only for lesser shear in the model than in the numerical simulation. The right panel of Fig.~\ref{fig1} illustrates that, in agreement with the DNS data, the convective heat flux is reduced with increasing shear. We stress the need for further numerical simulations to enable us to make more detailed comparisons over a much broader range of Rayleigh and Prandtl numbers. 

\begin{acknowledgement}
This research is supported by the Austrian FWF grant No. P21205-N16. Support from \"Osterreichische Forschungsgemeinschaft (Projekt 06/12308) is greatly acknowledged. DOG is grateful to the Leverhulme Trust for an Emeritus Fellowship. 
\end{acknowledgement}

\end{document}